\documentclass[11pt,letterpaper,leqno]{article}
\usepackage[T1]{fontenc}
\usepackage{lmodern}
\usepackage{amssymb,amsfonts,mathtools}
\usepackage{graphicx,booktabs,multirow}
\usepackage{subcaption,float,placeins,enumitem,bbm,url}
\usepackage[authoryear,round]{natbib}
\usepackage[textwidth=6.151in,textheight=594.8005pt,centering]{geometry}
\usepackage{amsthm,xcolor}
\usepackage[hidelinks]{hyperref}
\hypersetup{pdftitle={Testing Conditional Distribution Equality Using Generative Models},pdfauthor={Zekai Jin, Yi Zhang, Yun Yang, Xiaofeng Shao}}
\numberwithin{equation}{section}
\newtheorem{theorem}{Theorem}[section]
\newtheorem{lemma}[theorem]{Lemma}
\theoremstyle{remark}
\newtheorem{remark}[theorem]{Remark}
\newtheorem{assumption}[theorem]{Assumption}
\newenvironment{keywords}{\par\smallskip\noindent\textbf{Keywords.}\ }{\par}
\newenvironment{MSCcodes}{\par\smallskip\noindent\textbf{MSC codes.}\ }{\par\medskip}
\newcommand{\email}[1]{\href{mailto:#1}{\nolinkurl{#1}}}
\date{}

\newcommand{\bG}{\mathbb{G}}

\newcommand{\cK}{\mathcal{K}}
\newcommand{\cJ}{\mathcal{J}}
\newcommand{\cN}{\mathcal{N}}

\newcommand{\cH}{\mathcal{H}}
\newcommand{\R}{\mathbb{R}}
\newcommand{\wt}{\widetilde}
\newcommand{\wh}{\widehat}
\newcommand{\PP}{\mathbb{P}}

\newcommand{\E}{\mathbb{E}}

\DeclareMathOperator*{\var}{Var}

\newcommand{\indep}{\mathrel{\perp\!\!\!\perp}}
\newcommand{\Var}{\operatorname{Var}}

\title{Testing Conditional Distribution Equality Using Generative Models%
\thanks{Zekai Jin and Yi Zhang contributed equally to this work and are listed in alphabetical order.}}
\author{Zekai Jin\thanks{Department of Mathematics, University of Maryland, College Park,
College Park, MD 20742, USA (\email{zekaijin@umd.edu}).}
\and Yi Zhang\thanks{Department of Statistics and Data Science, Washington University in St Louis,
St Louis, MO 63130, USA (\email{yiz@wustl.edu}). Corresponding author.}
\and Yun Yang\thanks{Department of Mathematics, University of Maryland, College Park,
College Park, MD 20742, USA (\email{yy84@umd.edu}).}
\and Xiaofeng Shao\thanks{Department of Statistics and Data Science and Department of Economics,
Washington University in St Louis, St Louis, MO 63130, USA (\email{shaox@wustl.edu}).}}

\begin{document}
\maketitle
\begin{abstract}
		Conditional distribution equality testing is a core problem with applications in causal inference, model diagnostics, and transfer learning. Existing methods often lose power when the conditioning variable is high-dimensional. We propose a new test that leverages generative models to compare conditional distributions using both observed data and synthetic samples. A sample-splitting and studentization scheme is introduced to control estimation error. Under the null, the test statistic is asymptotically normal, and we establish local power against contiguous alternatives. Our theory under the null and fixed alternative requires only consistency of the learned conditional distribution, allowing for a broad class of generative models. Empirical results demonstrate accurate size control and strong power in high-dimensional settings compared to representative kernel-based conditional and joint two-sample tests.

	\end{abstract}
\begin{keywords}
		Conditional distribution; generative model; kernel mean embedding; sample splitting; two-sample test.
	\end{keywords}
\begin{MSCcodes}
62G10, 62H15
\end{MSCcodes}
\section{Introduction}\label{sec:intro}
	
	Conditional distribution equality (CDE) testing is a fundamental problem in modern statistics and machine learning, with applications in transfer learning \citep{maqsood2019transfer, zhuang2020comprehensive}, fairness auditing \citep{cherian2024statistical}, representation learning \citep{bengio2013representation, huang2024deep}, and simulator validation/ calibration \citep{degroot1983comparison,murphy1977reliability}. 
	In many contemporary pipelines, data are generated or evaluated conditional on high-dimensional contextual variables or learned representations \citep{zhao2021diagnostics,dey2022calibrated}, 
	making it essential to assess whether conditional laws coincide in order to validate distributional alignment and model robustness. Unlike marginal two-sample testing, CDE testing requires comparing entire conditional distributions across the support of the conditioning variable, which substantially increases statistical complexity.
	
	Specifically, Let $X,Y\in\R^{d_X}$ and $Z\in\R^{d_Z}$ be random vectors. We consider the problem of testing whether the conditional distributions of $X$ and $Y$ given $Z$ are equal, that is,
	\begin{align}
		H_0:\; \PP_{X\mid Z}=\PP_{Y\mid Z} ~\text{almost surely under \ } \PP_Z,
		\text{versus}  ~H_1:\; \PP_Z\big(\PP_{X\mid Z}\neq \PP_{Y\mid Z}\big)>0.\label{eq_hypothesis}
	\end{align}
	This is a paired conditional two-sample setting in which, for each covariate/context $Z_i$, one observes or can generate both an $X_i$
	-response and a $Y_
	i$-response, and the two responses may be dependent given $Z_i$. This formulation arises naturally in many statistical and machine learning problems. The leading example is the evaluation of a \emph{candidate} model: $X_i$ is the observed response and $Y_i$ is a draw at the same $Z_i$ from a fitted predictive or generative model (simulator) so that $H_0$ states that the candidate reproduces the target conditional distribution \citep{cranmer2020frontier,wood2010statistical}. The same structure applies when comparing conditional outputs of two models \citep{degroot1983comparison} and comparing regression curves \citep{hardle1990semiparametric}; see Section 1 of \citet{chatterjee2024ecmmd} for a detailed account of these motivating examples. Throughout, the candidate model, which only produces data and is never assumed to be accurate, should be distinguished from the \emph{auxiliary} conditional generators that we fit inside the testing procedure.
	
	Existing kernel- and distance-based CDE tests characterize conditional distribution equality through various population measures/discrepancies, such as integrated conditional maximum mean discrepancy or energy distance \citep{chatterjee2024ecmmd,yan2025cond2sample}. 
	While these methods avoid explicit conditional density estimation, which is difficult and computationally infeasible in the high dimensional setting, they usually rely on smoothing or constructing local neighborhoods (of the conditioning variable $Z$) to estimate a function/functional of $Z$. As $d_Z$ increases, distances in the ambient conditioning space become less informative and such procedures may suffer from bias inflation, unstable calibration, and loss of power, particularly when $Z$ represents moderate or large dimensional learned features from modern representation models; see the simulation results in Section~\ref{sec:simu} and the real data example in Section ~\ref{sec:data}. 
	
	Alternatively, the CDE testing problem can be viewed as a special case of the standard two-sample testing problem for the joint distributions of $(X, Z)$ and $(Y, Z)$, with the additional structure that the second component $Z$ is shared across the independent samples. However, directly applying standard two-sample testing procedures to CDE testing may lead to severe power loss, especially in the presence of a potentially high-dimensional conditioning variable $Z$. Heuristically, this occurs because a standard two-sample test does not account for the additional structure that the conditioning variable $Z$ is shared between the two joint distributions, while the signal appears only in the conditional distributions of $X$ and $Y$ given $Z$. Lemma \ref{lemma_mmd_high_dim} in Section \ref{sec:pop-measure} makes this precise for the maximum mean discrepancy (MMD) with a product kernel: as $d_Z\to\infty$, the MMD between the two joint distributions converges to a constant multiple of the MMD between the marginal distributions of $X$ and $Y$. A joint test is then blind to alternatives under which the conditional distributions differ but the marginals agree, as in Example~1 of Section~\ref{sec:simu}, where the joint MMD test has trivial power once $d_Z\ge 200$;; see also Remark \ref{rem:power-intuition} in Section \ref{sec:theory}.

	In this work, we adopt a different perspective by replacing kernel smoothing based estimation with sampling from learned conditional generators. Those conditional generators, usually learned using generative neural networks, may exploit sparse, low-dimensional or compositional structure even when the ambient dimension of $Z$ is large. Instead of estimating functions of $Z$, we construct Monte Carlo approximations based on samples drawn from estimated conditional distributions. To this end, we leverage recent advances in generative modeling 
	to learn conditional generators to sample from $\PP_{X\mid Z}$ and $\PP_{Y\mid Z}$. 
	Our approach is model-agnostic and can accommodate many conditional generative modeling methods, such as generative adversarial networks (GAN) \citep{goodfellow2014generative}, generative moment matching networks (GMMN) \citep{dziugaite2015training, li2015generative}, mixture density networks (MDN) \citep{bishop1994mixture, papamakarios2016fast}, Engression \citep{shen2025engression} and diffusion models \citep{NEURIPS2020_4c5bcfec}. More importantly, the generator estimation errors add no bias to our test statistic under the null, and only consistency of the generators is required, with no rate requirement on their accuracy, which allows the method to accommodate a wide range of commonly used generative models. We emphasize that consistency of the fitted generators is a very weak requirement on their accuracy: the estimation error need only vanish, at an arbitrarily slow rate. Under the regularity conditions of Section~\ref{sec:theory}, this suffices for asymptotic validity under the null and for power tending to one against fixed alternatives. To the best of our knowledge, this is the weakest requirement on generator accuracy among existing generative model-based tests of this kind; see Section~\ref{sec:related} and the discussion following Theorem~\ref{th_null}.

	Building on this framework, we propose a studentized two-split test statistic for $H_0$. Sample splitting separates generator training from test statistic evaluation, while studentization absorbs both Monte Carlo variability and estimation error arising from nuisance components. As a result, the proposed test statistic converges to a mean-zero normal distribution under the null, yielding closed-form critical values and eliminating the need for bootstrap resampling or permutation. Extensive numerical experiments demonstrate that the proposed procedure achieves accurate size control and competitive power relative to existing kernel- and energy-based CDE tests, particularly in high-dimensional settings. These results highlight the robustness and scalability of the conditional sampling approach for modern data regimes.
	
	\subsection{Related Literature and Our Contributions}\label{sec:related}

	Kernel methods provide a flexible framework for comparing distributions without the need for (conditional) density estimation. In particular, the maximum mean discrepancy \citep[MMD;][]{gretton2012mmd}
	and its variants have been widely used in marginal two-sample testing. Extending these ideas to the conditional setting, recent work introduces population-level discrepancies that integrate distances between conditional distributions over the distribution of the conditioning variable. Notable examples include the expected conditional MMD \citep[ECMMD;][]{HuangLamZhang2022Aleatoric,chatterjee2024ecmmd} and related integrated conditional energy distances \citep{yan2025cond2sample}, which vanish if and only if the conditional distributions coincide under suitable conditions. More recently, \citet{MoskvichevEtAl2026Kernel} studied a hierarchy of kernel-based conditional MMD measures, including the conditional-mean-operator version CMMD$_0$ used in our simulations. These approaches yield elegant theoretical characterizations and lead to U-statistic--type estimators.  Despite these advances, practical implementations of kernel- and distance-based conditional tests typically rely on smoothing, nearest-neighbor constructions \citep{chatterjee2024ecmmd}, or local aggregation in the space of the conditioning variable \citep{yan2025cond2sample}. Consequently, their finite-sample performance may deteriorate when the conditioning variable is high-dimensional due to the curse of dimensionality; see the numerical evidence in Section ~\ref{sec:simu}.
	
	It is also worth noting that a related but more general testing problem,
	$H_0:\,\PP_{Y_1\mid X_1}=\PP_{Y_2\mid X_2}$, has been studied in \citet{HuLei2023,yan2025cond2sample} based on independent samples $\{(X_{1i},Y_{1i})\}_{i=1}^{n_1}$ and $\{(X_{2j},Y_{2j})\}_{j=1}^{n_2}$. In \citet{HuLei2023}, the conditioning variables $X_1$ and $X_2$ are allowed to have different distributions, provided that they share the same support and are independent. Our framework differs in that the conditioning variable $Z$ is shared across samples; a similar setting is also considered in \citet{chatterjee2024ecmmd}. 
	As a result, the methods from \citet{HuLei2023} are not directly applicable to our setting. Moreover, their conformal prediction-based approach requires estimation of both marginal and conditional density ratios via classification, which can be challenging when the conditioning variable is high-dimensional. Recently, \citet{GaoHuangYangShao2026} addressed this more general problem with conditional generative models, without requiring the two samples to be independent; in exchange, their theory imposes rate conditions on the generator estimation errors and requires the two responses to be conditionally independent given the conditioning variables.

	More recently, generative modeling techniques have been explored to facilitate statistical inference involving conditional distributions in a variety of settings, such as conditional independence testing \citep{zhang2025drci,ShiXuBergsmaLi2021DoubleGANsCI,he2025conditional}, variable selection \citep{romano2020deep}, feature importance evaluation \citep{blesch2025conditional} and causal inference \citep{chen2025enhancing}. For example, \citet{zhang2025drci} proposed a doubly robust conditional independence testing procedure based on learned conditional distributions, extending earlier work by \citet{ShiXuBergsmaLi2021DoubleGANsCI}. Both of these approaches impose explicit rate conditions on the estimation error of the learned conditional distributions, namely that the product of two generator errors be $o(n^{-1/2})$, whereas our theoretical analysis requires only consistency under the null and fixed alternatives. 
	
	Our contributions build on these developments. We introduce a conditional sampling-based framework for CDE testing that approximates the population discrepancy using Monte Carlo samples from learned conditional distributions, without local neighborhoods or kernel smoothing in the conditioning space. Sample splitting and studentization yield an asymptotically normal test statistic under the null, with closed-form critical values that avoid bootstrap or permutation calibration.
	
	Under the stated regularity conditions, our theory requires only consistency of the learned conditional distributions under the null and fixed alternatives, allowing a broad class of conditional generative models to be used. We also establish nontrivial local power against alternatives whose conditional mean embeddings differ at the $n^{-1/4}$ rate, subject to the corresponding generator-estimation conditions. Such a local asymptotic power analysis was not developed in \citet{chatterjee2024ecmmd}. Finally, extensive simulations and a real data application demonstrate accurate size control and competitive power relative to kernel-based conditional and joint two-sample tests, particularly when the conditioning variable is high-dimensional or consists of learned representations.

	\subsection{Notation}
	We now introduce notation used throughout the paper.
	For any positive integer $d$ and random vectors $(X^1,X^2,\dots,X^d,Z)$ defined on the same probability space, we use $\PP_{X^1X^2\cdots X^d}$ and $\PP_{X^1X^2\cdots X^d\mid Z}$ to denote the joint distribution of $(X^1,X^2,\dots,X^d)$ and its conditional distribution given $Z$.
	Let $\E_Z$ or $\E_{Z\sim Q}$ denote expectation with respect to a random variable $Z$ with distribution $Q$, and write $\E$ when there is no ambiguity.
	For a positive integer $n$, let $[n]:=\{1,2,\ldots,n\}$.
	Let $\mathcal H$ denote a separable reproducing kernel Hilbert space (RKHS) associated with a positive definite kernel $\cK(\cdot,\cdot)$.
	We write $\langle\cdot,\cdot\rangle_{\mathcal H}$ and $\|\cdot\|_{\mathcal H}$ for the inner product and induced norm on $\mathcal H$, respectively.
	For two Hilbert spaces $\mathcal H_1$ and $\mathcal H_2$, their tensor product is denoted by $\mathcal H_1\otimes\mathcal H_2$.
	Throughout the paper, we consider kernels defined on Euclidean spaces; in particular, the Gaussian kernel is $\cK(a,b)=\exp\!\big(-\|a-b\|_2^2/(2\gamma^2)\big)$ and the Laplacian kernel is $\cK(a,b)=\exp\!\big(-\|a-b\|_1/\gamma\big)$, where $\gamma>0$ is a bandwidth parameter, $\|\cdot\|_2$ denotes the Euclidean norm, and $\|\cdot\|_1$ denotes the $\ell_1$ norm.
	Finally, we use $\xrightarrow{d}$ and $\xrightarrow{p}$ to denote convergence in distribution and convergence in probability, respectively.
	All asymptotic statements are taken as the sample size $n\to\infty$ unless stated otherwise.

	\section{Conditional Distribution Equality Testing via Conditional Sampling}
	\label{sec:method}
	
	In this section, we first introduce a population-level measure for evaluating conditional distribution equality and propose a corresponding sample-based statistic for CDE testing. We then establish theoretical guarantees for the proposed procedure, including size control under the null hypothesis and power against local alternatives.

	\subsection{Population CDE measure}\label{sec:pop-measure}
	
	Our population CDE measure is based on the maximum mean discrepancy (MMD), a widely used metric for quantifying discrepancies between probability distributions due to its strong theoretical guarantees, computational simplicity, and compatibility with kernel-based methods.
	Concretely, let $\cK_d:\R^{d}\times\R^{d}\to\R$ be a symmetric, positive-definite kernel on $\R^d$ with associated RKHS $\cH_d$ \citep{gretton2012mmd}. The squared MMD between two distributions $P$ and $Q$ on $\R^d$ is defined as $\text{MMD}^2(P,Q)=\|\mu_P-\mu_Q\|_{\cH_d}^2\ge 0$, where $\mu_P=\E_{X\sim P}\big[\cK_d(X,\cdot)\big]\in\cH_d$ denotes the kernel mean embedding of $P$ (and $\mu_Q$ is defined analogously). If $\cK_d$ is characteristic \citep[][Section~2.2]{FukumizuEtAl2007KCDC}, then $\text{MMD}^2(P,Q)=0$ if and only if $P=Q$. 
	
	Moreover, given random samples $\{X_i\}_{i=1}^n$ drawn i.i.d.\ from $P$ and $\{Y_j\}_{j=1}^m$ drawn i.i.d.\ from $Q$, $\text{MMD}^2(P,Q)$ admits an unbiased empirical estimator of the form
	\begin{align}
		\frac{1}{n(n-1)}\sum_{i\ne i'} \cK_d(X_i,X_{i'})
		+
		\frac{1}{m(m-1)}\sum_{j\ne j'} \cK_d(Y_j,Y_{j'})-
		\frac{2}{nm}\sum_{i,j} \cK_d(X_i,Y_j),\label{eqn:MMD_sample}
	\end{align}
	which makes it straightforward to compute in practice and also convenient for theoretical analysis due to its simple U-statistic structure.
	
	Motivated by related population measures for comparing conditional distributions in the literature (e.g., \citet{HuangLamZhang2022Aleatoric}), a natural idea is to construct a population CDE measure by directly comparing $\PP_{X\mid Z=z}$ and $\PP_{Y\mid Z=z}$ for every $z\in\R^{d_Z}$ in the support of $\PP_Z$, or equivalently by comparing the joint distributions $\PP_{XZ}$ and $\PP_{YZ}$. Concretely, one may consider either $\E_Z\big[\text{MMD}^2(\PP_{X\mid Z},\PP_{Y\mid Z})\big]$ or $\text{MMD}^2(\PP_{XZ},\PP_{YZ})$ as candidate population measures, since both vanish if and only if the null hypothesis in~\eqref{eq_hypothesis} holds. However, as we will show below, using $\text{MMD}^2(\PP_{XZ},\PP_{YZ})$ can suffer from substantial power loss when the dimension of $Z$ is large.

	To formally demonstrate the loss of power, we consider an asymptotic regime in which the distributions of $X$ and $Y$ remain fixed while the dimension $d_Z$ of $Z$ diverges to infinity. Let $\cK_{d_X}$ be a bounded kernel defined on $\R^{d_X}$, and let $\cK_Z(a,b)=f(\|a-b\|/\gamma)$ be a kernel on $\R^{d_Z}$ for some bounded  and continuous function $f$ and bandwidth $\gamma>0$. Define the product kernel $\cK_d=\cK_{d_X}\times \cK_Z$. The following lemma (whose proof is provided in Section SM2 of the Supplementary Material) shows that under certain conditions, the MMD between the two joint distributions $\PP_{XZ}$ and $\PP_{YZ}$ converges, up to a constant factor, to the MMD between the two marginal distributions $\PP_X$ and $\PP_Y$.

	\begin{lemma}\label{lemma_mmd_high_dim}
		If $\|Z-Z'\|/\gamma \stackrel{p}{\to} a>0$ as $d_Z\to\infty$, where $Z,Z'\stackrel{iid}{\sim} \PP_Z$, then as $d_Z\to\infty$, we have
		\begin{equation*}
			\text{MMD}^2(\PP_{XZ},\PP_{YZ}) \stackrel{d_Z\to\infty}{\to} f(a)\,\text{MMD}^2(\PP_{X},\PP_{Y}).
		\end{equation*}
	\end{lemma}
	
	The assumption $\|Z-Z'\|/\gamma \stackrel{p}{\to} a>0$ as $d_Z\to\infty$ in the lemma is satisfied, for example, when the median heuristic is used to choose $\gamma$ as the median of $\|Z-Z'\|$, the variance of each component of $Z$ is uniformly bounded away from zero, and the second moment of each component is uniformly bounded above; see Assumption D1 in \citet{zhu2020distance} for details. Intuitively, this condition suggests that the kernel $\cK_Z$ becomes nearly constant in high dimensions, so that differences in the $Z$ component are no longer informative in the MMD computation, and the MMD effectively reduces to comparing the marginal distributions.

	This result indicates that as the dimension $d_Z$ diverges, CDE testing procedures based on $\text{MMD}^2(\PP_{XZ},\PP_{YZ})$ (e.g., the MMD two-sample test in \citet{gretton2012mmd}) become asymptotically equivalent to testing the equality of the marginal distributions $\PP_X$ and $\PP_Y$. Consequently, when $Z$ is high-dimensional, such procedures may fail to reject $H_0$ and lose power in situations where the conditional distributions $\PP_{X\mid Z}$ and $\PP_{Y\mid Z}$ differ but the marginals $\PP_X$ and $\PP_Y$ remain similar.

	Motivated by this collapse issue of $\text{MMD}^2(\PP_{XZ},\PP_{YZ})$ when the conditioning variable $Z$ is high-dimensional, we adopt $\E_Z\big[\text{MMD}^2(\PP_{X\mid Z},\PP_{Y\mid Z})\big]$ as our population CDE measure, as also considered in \citet{chatterjee2024ecmmd}. Specifically, let $\cK=\cK_{d_X}:\R^{d_X}\times\R^{d_X}\to\R$ be a characteristic, positive-definite kernel inducing an RKHS $\cH_0$. The resulting population CDE measure, referred to as the expected conditional maximum mean discrepancy (ECMMD), is defined as
	\begin{align}
		{\rm ECMMD}^2\big[\PP_{X\mid Z},\PP_{Y\mid Z}\big]
		=\,& \E\Big[\,\big\|\,\E\big[\cK(X,\cdot)\mid Z\big]\,-\,\E\big[\cK(Y,\cdot)\mid Z\big]\,\big\|^2_{\cH_0}\,\Big]\nonumber\\ 
		=\,&\E\big[\,\cK(X,X')+\cK(Y,Y')-\cK(X,Y')-\cK(X',Y)\,\big].\notag
	\end{align}
	Here, the expectation on the second line is taken with respect to $(X,X')$ and $(Y,Y')$ such that
	$\PP_{X|Z}=\PP_{X'|Z}$, $\PP_{Y|Z}=\PP_{Y'|Z}$, and $(X,Y)\indep (X',Y')\mid Z$. It can be shown that ${\rm ECMMD}^2=0$ if and only if $\PP_{X\mid Z}=\PP_{Y\mid Z}$ almost surely with respect to $\PP_Z$; see Proposition~2.4 of \citet{chatterjee2024ecmmd}. When $\cK(u,v)=-\|u-v\|$, ${\rm ECMMD}^2$ reduces to the expected conditional energy distance proposed in \citet{yan2025cond2sample}. This representation also helps explain why ${\rm ECMMD}^2$ is less sensitive to high-dimensional $Z$, since $Z$ does not appear explicitly in the sample version of the statistic, but only implicitly through expectations over $(X,Y)$ sampled under the same $Z$.
	
	To incorporate conditional generative models and emphasize the role of $Z$, we introduce an alternative representation of ${\rm ECMMD}^2$:
	\begin{align}
		{\rm ECMMD}^2\big(\PP_{X\mid Z},\PP_{Y\mid Z}\big)
		=
		\E\Big[\,\big\langle \cK(X,\,\cdot\,)-\cK(Y,\,\cdot\,),\, g_X(Z)-g_Y(Z)\big\rangle_{\cH_0}\Big],
	\end{align}
	where $g_X(z)=\E[\,\cK(X,\,\cdot\,)\mid Z=z]$ and $g_Y(z)=\E[\,\cK(Y,\,\cdot\,)\mid Z=z]$ denote the respective kernel conditional mean embeddings of $\PP_{X\mid Z}$ and $\PP_{Y\mid Z}$. Our test statistic is constructed from a sample analogue of this representation, similar to equation~\eqref{eqn:MMD_sample}.

	\subsection{Testing procedure}\label{sec:procedure}
	
	To construct a feasible sample version of ${\rm ECMMD}^2$, we estimate the conditional mean embeddings $g_X(\cdot)$ and $g_Y(\cdot)$. Classical nonparametric methods (e.g., kernel smoothing or nearest-neighbor regression) become inaccurate and computationally prohibitive when the conditioning variable $Z$ is high-dimensional. Instead, we adopt a conditional \emph{sampling-based} strategy, replacing smoothing with Monte Carlo sampling from learned conditional generators.
	
	\smallskip
	\noindent\textit{Conditional generators.}
	We fit two generative models to sample from $\PP_{X\mid Z}$ and $\PP_{Y\mid Z}$, respectively. Specifically, by the noise-outsourcing lemma (Theorem 6.10 of \citet{kallenberg2002foundations}; Lemma 2.1 of \citet{zhou2022deep}), for any integer $d_N \ge 1$, there exists a measurable function $\bG_X$ such that, for $\kappa \sim \cN(0, I_{d_N})$ independent of $Z$, we have $\bG_X(\kappa, Z)\mid Z \sim \PP_{X\mid Z}$ (and analogously for $\bG_Y$). Based on this result, we train a generative neural network to approximate $\bG_X$; see Section SM1 of the Supplementary Material for details.

	In principle, any conditional generative modeling procedure proposed in the literature may be used.
	In this work, we adopt the conditional diffusion model, Engression, and MDN in Sections \ref{sec:simu} and \ref{sec:data} to train two neural networks $\wh{\bG}_X:\R^{d_Z}\times\R^{d_N}\to\R^{d_X}$ and $\wh{\bG}_Y:\R^{d_Z}\times\R^{d_N}\to\R^{d_X}$, such that $\wh{\bG}_X(Z,\kappa)$ and $\wh{\bG}_Y(Z,\eta)$ can be viewed as samples from approximations of $\PP_{X\mid Z}$ and $\PP_{Y\mid Z}$, respectively.
	Here $\kappa,\eta\in\R^{d_N}$ are independent noise variables, independent of $(X,Y,Z)$, and drawn from a known reference distribution, such as a standard multivariate normal distribution or a uniform distribution on the unit hypercube. The noise dimension $d_N$ and the reference distribution depend on the chosen generative modeling procedure.
	In the simulation and real data examples in Sections \ref{sec:simu} and \ref{sec:data}, for Engression, we set $d_N=50$ and take $\kappa,\eta\sim \mathrm{Unif}(0,1)^{50}$, following the practice in \citet{shen2025engression}. 
	For MDN and the conditional diffusion model, we set $\kappa,\eta\sim \cN(0,1)$.
	
	\smallskip
	\noindent\textit{Sample splitting and cross-fitting.}
	To preserve independence between training and testing, we adopt a sample-splitting and cross-fitting strategy. Specifically, for positive integers $n$ and $M$, let $\{(X_i,Y_i,Z_i)\}_{i=1}^{2n}$ be i.i.d.~copies of $(X,Y,Z)$, and let $\{\kappa_i^m,\eta_i^m\}_{i=1,\dots,2n;\,m=1,\dots,M}$ be i.i.d.~copies of $(\kappa,\eta)$, independent of the data. 
	We partition $[2n]$ into two equal folds, $\cJ^{(1)}=\{1,\dots,n\}$ and $\cJ^{(2)}=\{n+1,\dots,2n\}$, and write $\cJ^{(-j)}=[2n]\setminus\cJ^{(j)}$ for $j\in[2]$. For each fold $j$, we train generators $\wh\bG_{Xj}$ and $\wh\bG_{Yj}$ using data $\{(X_i,Z_i)\}_{i\in\cJ^{(-j)}}$ and $\{(Y_i,Z_i)\}_{i\in\cJ^{(-j)}}$, respectively. For $i\in\cJ^{(j)}$, define $\wh X_i^{(m)}=\wh\bG_{Xj}(Z_i,\kappa_i^m)$ and $\wh Y_i^{(m)}=\wh\bG_{Yj}(Z_i,\eta_i^m)$, and set $\wh g_X(Z_i)=M^{-1}\sum_{m=1}^M \cK(\wh X_i^{(m)},\cdot)$ and $\wh g_Y(Z_i)=M^{-1}\sum_{m=1}^M \cK(\wh Y_i^{(m)},\cdot)$.
	
	On each fold $\cJ^{(j)}$, we define the studentized statistic $\wh T_n^{(j)}=n^{-1/2}\sum_{i\in\cJ^{(j)}} \wh S_i/\wh\sigma_n^{(j)}$, where
	\begin{equation}\label{eq_defs1}
	\begin{aligned}
		\wh S_i
		&=
		\Big\langle \cK(X_i,\cdot)-\cK(Y_i,\cdot),\,\wh g_X(Z_i)-\wh g_Y(Z_i)\Big\rangle_{\cH_0} \\
		&=
		\frac{1}{M}\sum_{m=1}^M \Big\{\cK(X_i,\wh X_i^{(m)})+\cK(Y_i,\wh Y_i^{(m)}) -\cK(X_i,\wh Y_i^{(m)})-\cK(\wh X_i^{(m)},Y_i)\Big\},
	\end{aligned}
	\end{equation}
	and $\wh\sigma_n^{(j)}$ denotes the sample standard deviation of $\{\wh S_i\}_{i\in\cJ^{(j)}}$. The final test statistic is $\wh T_n=(\wh T_n^{(1)}+\wh T_n^{(2)})/2$. According to Theorem~\ref{th_null}, $\wh T_n$ converges to $T^\dagger\sim \cN(0,1/2)$ under $H_0$. Let $\xi_\tau$ be the $100(1-\tau)$-th percentile of $T^\dagger$ for any $\tau\in(0,1)$, our test will reject $H_0$ at level $\tau$ if $\wh T_n > \xi_\tau$.
	
	\begin{remark}
		Although both our test and the test proposed in Section 4 of \citet{chatterjee2024ecmmd} are based on the same population CDE measure, our procedure exhibits much better finite-sample performance when the dimension of $Z$ is large. This difference can be attributed to several factors. First, in \citet{chatterjee2024ecmmd}, the conditional mean embeddings $g_X(Z_i)$ and $g_Y(Z_i)$ are estimated by sample averages over a $K$-nearest neighbor graph of $Z_i$, which can be unreliable in high dimensions; see Table~\ref{tab_exp1} in Section~\ref{sec:simu}. Second, since their method does not use sample splitting or cross-fitting, strong dependence arises between the estimators of $g_X(Z_i)$ and $g_X(Z_j)$ for $i\neq j$, which may lead to size distortion under $H_0$; see Table~\ref{tab_exp2} in Section~\ref{sec:simu}. Third, as shown in Equation (4.1) of \citet{chatterjee2024ecmmd}, their test statistic depends explicitly on the choice of the number of nearest neighbors $K$, which can be difficult to tune in practice and may lead to different testing outcomes; see Figure~\ref{fig:combined} (a) and (b) in Section~\ref{sec:data}.
		
	\end{remark}

	\subsection{Theoretical properties}\label{sec:theory}
	
	In this subsection, we study the asymptotic properties of the proposed conditional two-sample test as the sample size $n$ diverges. Our analysis focuses on two key objectives. First, we show that the test attains asymptotically correct size, with the rejection probability converging to the nominal significance level. Second, we examine its asymptotic power against sequences of local alternatives approaching the null at polynomial rates.

	For $m\in[M]$, define the pseudo-samples $\wt X_i^{(m)}=\bG_X(Z_i,\kappa_i^m)$ and $\wt Y_i^{(m)}=\bG_Y(Z_i,\eta_i^m)$.
	Denote $\wt g_X(Z_i)=M^{-1}\sum_{m=1}^M \cK(\wt X_i^{(m)},\cdot)$, $\wt g_Y(Z_i)=M^{-1}\sum_{m=1}^M \cK(\wt Y_i^{(m)},\cdot)$, and define
	\begin{equation*}
		\wt S_i=\big\langle \cK(X_i,\cdot)-\cK(Y_i,\cdot),\,\wt g_X(Z_i)-\wt g_Y(Z_i)\big\rangle_{\cH_0}.
	\end{equation*}
	For $j\in[2]$, define the oracle statistic on each fold as
	$T_n^{(j)}=n^{-1/2}\sum_{i\in\cJ^{(j)}} \wt S_i/\sigma_n^{(j)}$, where $\sigma_n^{(j)}$ is the sample standard deviation of $\{\wt S_i\}_{i\in\cJ^{(j)}}$. The oracle statistic uses the true (unknown) generators and converges in distribution to $\cN(0,1)$ under $H_0$.
	The oracle analogue of $\wh T_n$ is $T_n=(T_n^{(1)}+T_n^{(2)})/2$.
	The following assumption is used to derive the asymptotic properties of the statistic.
	
	\begin{assumption}\label{assump2}
		Let $D_i\in\{X_i,\wt X_i^{(1)},\wh X_i^{(1)}\}$, $E_i\in\{Y_i,\wt Y_i^{(1)},\wh Y_i^{(1)}\}$, and $i\in\{i_1,i_2\}$ with $i_s\in\cJ^{(s)}$ for $s\in[2]$. There exist constants $C_0,C_1>0$ and $\beta\ge 0$, such that 
		\begin{enumerate}[label=(\alph*),leftmargin=2em,labelsep=0.6em]
			\item $\E\!\left[\{\cK(D_i,D_i)\}^3+\{\cK(E_i,E_i)\}^3\right]<C_0$.
			\item\label{assump2_b}$\begin{aligned}[t]
			&\E\big[(\cK(X_i,X_i)+\cK(Y_i,Y_i))
			\|\cK(\wt X_i^{(1)},\cdot)-\cK(\wh X_i^{(1)},\cdot)\|_{\cH_0}^2\big]
			=o(n^{-2\beta}),\\
			&\E\big[(\cK(X_i,X_i)+\cK(Y_i,Y_i))
			\|\cK(\wt Y_i^{(1)},\cdot)-\cK(\wh Y_i^{(1)},\cdot)\|_{\cH_0}^2\big]
			=o(n^{-2\beta}).
			\end{aligned}$
			\item $\Var(\wt S_i)>C_1>0$.
		\end{enumerate}
	\end{assumption}
	\noindent Part (a) ensures that the (conditional) RKHS mean embeddings are well defined and that a central limit theorem for Hilbert-space–valued random elements applies; it is satisfied by bounded kernels such as the Gaussian and Laplacian kernels. Part (b) requires the estimation errors of the conditional generators to decay at rate $n^{-\beta}$ for some $\beta\ge 0$. 
	Part (c) excludes degeneracy of the oracle summand, for example when $\cK(X,\cdot)-\cK(Y,\cdot)$ is almost surely zero or when the Monte Carlo embedding difference has vanishing variance.
	If $\cK$ is bounded, part (a) holds trivially and part (b) reduces to $\E	\|\cK(\wt X_i^{(1)},\cdot)-\cK(\wh X_i^{(1)},\cdot)\|_{\cH_0}^2=o(n^{-2\beta})$, and
	$\E	\|\cK(\wt Y_i^{(1)},\cdot)-\cK(\wh Y_i^{(1)},\cdot)\|_{\cH_0}^2=o(n^{-2\beta})$.
	The following theorem characterizes the asymptotic behavior of the proposed test statistic under the null hypothesis; its proof is given in Section SM3 of the Supplementary Material. Throughout the asymptotic analysis, $M$ is fixed.
	
	\begin{theorem}\label{th_null}
		Suppose Assumption~\ref{assump2} holds  with some $\beta\geq0$. Then, under $H_0$, we have $\wh T_n-T_n=o_p(n^{-\beta})$ as $n\to\infty$. Moreover, $\wh T_n \xrightarrow{d} T^\dagger=(T_1^\dagger+T_2^\dagger)/2$, where $T_1^\dagger$ and $T_2^\dagger$ are independent standard normal random variables.
	\end{theorem}
	
	\noindent \textcolor{black}{Compared with existing generative model-based nonparametric tests such as \citet{ShiXuBergsmaLi2021DoubleGANsCI} and \citet{zhang2025drci} which requires the product of the estimation errors in $\wh \bG_X$ and $\wh \bG_Y$ to decay faster than $n^{-1/2}$ (see Theorem 4 in \citet{ShiXuBergsmaLi2021DoubleGANsCI} and Assumption 2 in \citet{zhang2025drci}), Theorem \ref{th_null} only requires the consistency of the generative model estimators with no explicit rate conditions. }Our asymptotic theory is conditional on successful estimation of the conditional mean embeddings/generators. We do not prove consistency of a particular neural network based generator in arbitrary high dimension. Rather, the result shows that once such a generator is consistent, no additional smoothing over $Z$ is needed by the test statistic.

	To study the asymptotic power of the test, we consider a sequence of local alternatives under which the discrepancy between the conditional distributions vanishes as the sample size increases. Let $(X^0,Y^0,Z^0)$ denote the random triple $(X,Y,Z)$ under $H_0$, and consider a sequence of alternatives $(X_n^A,Y^0,Z^0)$ under which
	\begin{align}\label{eq_assump_power}
		H_{1n}:
		\E\!\big[\cK(X_n^A,\cdot)\mid Z^0\big]
		=
		\E\!\big[\cK(Y^0,\cdot)\mid Z^0\big]
		+
		n^{-\alpha}\Delta(Z^0),
	\end{align}
	where $\Delta:\R^{d_Z}\to\cH_0$ is an $\cH_0$-valued function independent of $n$ and satisfies $\E\|\Delta(Z^0)\|_{\cH_0}^2=c>0$. Under $H_{1n}$, \textcolor{black}{the distribution of $(Y,Z)$ does not change with $n$ and is the same as that under $H_0$. In addition, ${\rm ECMMD}^2\!\big(\PP_{X^A_n\mid Z^0},\PP_{Y^0\mid Z^0}\big) = \E[\wt S_i]=n^{-2\alpha}c$ so that the exponent $\alpha \geq 0$ determines the decay rate of the deviation from the null under $H_{1n}$; for instance, setting $\alpha = 0$ corresponds to a fixed alternative.} 
	The following theorem characterizes the asymptotic behavior of the proposed test statistic under $H_{1n}$; its proof is provided in Section SM4 of the Supplementary Material.

	\begin{theorem}\label{th_power}
		Suppose Assumptions~\ref{assump2} holds with $\beta\geq \alpha$, then we have that under $H_{1n}$:
		\begin{equation*}
		\begin{aligned}
			T_n-\frac{n^{1/2-2\alpha}c}{\sqrt{\var(\wt S_1)}}
			&\stackrel{d}{\to} T^\dagger, \\
			\wh T_n-T_n
			&=o_p\!\left(\max\{n^{1/2-2\alpha},1\}\right),
		\end{aligned}
		\end{equation*}
		which implies that: 1, If $0\leq \alpha<1/4$, then $\wh T_n \stackrel{p}{\to} \infty$. 2, If $\alpha=1/4$, then $\wh T_n- \frac{c}{\sqrt{\var(\wt S_1)}} \stackrel{d}{\to} T^\dagger$. 3, If $\alpha>1/4$, then $\wh T_n \stackrel{d}{\to} T^\dagger$.
		
	\end{theorem}
	
	\begin{remark}\label{rem:power-intuition}
		
		The first assertion in the above theorem shows that when $\alpha < 1/4$, the test statistic diverges to infinity, ensuring that the probability of our test not rejecting $H_0$ (i.e., Type II error) converges to zero. When $\alpha > 1/4$, the test statistic converges to the same limiting distribution as under the null hypothesis and the probability of Type II error will converge to $1-\tau$.
		
		The threshold $\alpha = 1/4$ (leading to an $n^{-1/4}$ rate) in our method differs from the $n^{-1/2}$ rate achieved by the MMD-based two-sample test in \citet{gretton2012mmd}. This slower rate reflects the price of robustness to the dimensionality of $Z$. Specifically, our statistic is computed from $\mathbb{R}^{d_X}$-valued variables $(X_i, Y_i, \widehat X_i^{(m)}, \widehat Y_i^{(m)})$, while the conditioning variable $Z \in \mathbb{R}^{d_Z}$ is used only to estimate the generators $\widehat{\mathbf{G}}_X$ and $\widehat{\mathbf{G}}_Y$. In contrast, \citet{gretton2012mmd} computes the statistic using $\mathbb{R}^{d_X + d_Z}$-valued variables $(X_i, Z_i)$ and $(Y_i, Z_i)$, requiring $Z$ to be fixed and low-dimensional. Our method remains valid even when $d_Z$ is high or grows with $n$, provided the conditional generators are consistent. As shown in Section~\ref{sec:simu}, the MMD-based test suffers significant power loss when $d_Z$ is large.

		Note that the required rate $o(n^{-\beta})$ for the generative-model estimation error depends on the decay rate of the signal. This condition guarantees that the plug-in error is asymptotically dominated by the signal magnitude; see the proof of Theorem~\ref{th_power} in Section SM4 of the Supplementary Material. When $\beta < \alpha$, the discrepancy between $\widehat T_n$ and $T_n$ is no longer negligible. In this regime, establishing the asymptotic behavior of $\widehat T_n$ would require an explicit characterization of the estimator's error, which is typically unavailable for black-box estimators such as neural networks. Under fixed alternative ($\alpha=0$), we only require the neural network estimators to be consistent in order for our test to have asymptotic power one.

	\end{remark}

	\section{Simulation Results}\label{sec:simu}

	To evaluate the finite-sample performance of our test, we consider two examples. In the first example, $X$ and $Y$ are univariate while the dimension of $Z$ can be large. In the second example, $X$ and $Y$ are multivariate with high-dimensional $Z$. Each experiment is repeated 500 times with sample size $n=800$ and nominal level 5\%;  see Section SM1 of the Supplementary Material for more implementation details.
	
	{\textbf{Example 1:}} \textit{
		Consider the model
		$X_i = m(Z_i) + \varepsilon_{Xi}$ and
		$Y_i = 2(\delta_i-1/2) m(Z_i) + \varepsilon_{Yi}$ for $i\in[n]$,
		where $Z_i =(Z_{1i},Z_{2i},\ldots,Z_{pi})^\top\stackrel{iid}{\sim} \cN(0,\Sigma_p)$ and the covariance matrix $\Sigma_p$ has entries $\Sigma_{ij} = (-0.3)^{|i-j|}$ for $i,j\in[p]$.
		Let $\delta_i \stackrel{i.i.d.}{\sim} \text{Bernoulli}(1-p_{flip})$, $\varepsilon_{Xi},\varepsilon_{Yi}\stackrel{iid}{\sim} \cN(0,1)$, and assume all variables are mutually independent.
		For $p\in\{1,10,200,2000\}$,
		$m(Z_i) = Z_{1i}$ if $p=1$ and $m(Z_i) = Z_{1i}+Z_{2i}$ otherwise.}
	
	{\textbf{Example 2:}} \textit{
		Fix $p=200$ and consider the model
		$X_i = m_s(Z_i) + \varepsilon_{Xi}$ and
		$Y_i = 2(\delta_i-1/2) m_s(Z_i) + \varepsilon_{Yi}$ for $i\in[n]$,
		where $(Z_i,\delta_i)$ are the same as in Example 1.
		For $s\in\{7,14,21\}$, let $\varepsilon_{Xi},\varepsilon_{Yi}\stackrel{iid}{\sim} \cN(0,I_s)$ and define
		$m_s(Z_i) = (Z_{1i},Z_{2i},\ldots,Z_{si})^\top$.}
	
	Here $p_{flip}=0$ corresponds to $H_0$, and the signal under $H_{1n}$ increases as $p_{flip}$ ranges over $\{0.2,0.4,0.6,0.8\}$. In Example 1 the conditional signal depends on only one or two coordinates of $Z$, so large $p$ adds many irrelevant conditioning coordinates; in Example 2 it depends on the first $s$ coordinates. For our method (denoted as GM$_1$), we set the number of synthetic samples to $M=1$, as additional simulations suggest the performance is not sensitive to $M$.
	For comparison, we include the test proposed in \citet{chatterjee2024ecmmd} (denoted NN$_{15}$ with 15 nearest neighbors). We also tried NN$_{25}$, following the recommendation in their paper, but do not present it because the results are similar. We include the permutation-based two-sample test on $\{(X_i,Z_i)\}_{i=1}^{400}$ and $\{(Y_i,Z_i)\}_{i=401}^{800}$ from \citet{gretton2012mmd} (denoted MMD). For Example~1, we additionally report the conditional test of \citet{yan2025cond2sample} (YLZ) and the CMMD$_0$ permutation test of \citet{MoskvichevEtAl2026Kernel} (MCS). YLZ estimates the integrated conditional discrepancy via conditional U-statistics and multivariate kernel smoothing in $Z$; its bandwidth conditions depend explicitly on $d_Z$, so it may suffer power loss as $d_Z$ increases. MCS estimates the conditional discrepancy via the Hilbert–Schmidt distance between kernel-ridge estimates of the two conditional mean operators; it avoids explicit kernel smoothing in $Z$ but requires regularized inversion of the covariate Gram matrices, and its power may deteriorate when kernel similarities become uninformative in high dimensions. Both tests are formulated for two independent samples, so we apply them to independent halves of the sample, as with the MMD baseline. implementation details are in Section SM1.1 of the Supplementary Material.

	\begin{table}[!htbp]
		\centering
		\caption{Empirical rejection rates (\%) for Example 1. The largest power, including ties, is in bold.}
		\label{tab_exp1}
		\small
		\setlength{\tabcolsep}{12pt}
		\renewcommand{\arraystretch}{0.82}
		\begin{tabular}{ccrrrrr}
			\toprule
			$p$ & $p_{\mathrm{flip}}$ & GM$_1$ & NN$_{15}$ & MMD & YLZ & MCS \\
			\midrule
			\multirow{5}{*}{1} & 0 & 6.0 & 3.4 & 4.2 & 5.8 & 4.2 \\
			& 0.2 & 16.4 & 44.2 & 37.8 & 18.4 & \textbf{88.6} \\
			& 0.4 & 61.4 & \textbf{100.0} & 99.8 & 84.0 & \textbf{100.0} \\
			& 0.6 & 95.8 & \textbf{100.0} & \textbf{100.0} & 99.8 & \textbf{100.0} \\
			& 0.8 & \textbf{100.0} & \textbf{100.0} & \textbf{100.0} & \textbf{100.0} & \textbf{100.0} \\
			\midrule
			\multirow{5}{*}{10} & 0 & 3.6 & 6.0 & 4.6 & 0.2 & 5.2 \\
			& 0.2 & 18.4 & 23.6 & 8.2 & 0.4 & \textbf{60.0} \\
			& 0.4 & 71.0 & 99.0 & 28.4 & 0.4 & \textbf{100.0} \\
			& 0.6 & 99.2 & \textbf{100.0} & 91.6 & 1.4 & \textbf{100.0} \\
			& 0.8 & \textbf{100.0} & \textbf{100.0} & \textbf{100.0} & 1.8 & \textbf{100.0} \\
			\midrule
			\multirow{5}{*}{200} & 0 & 5.0 & 5.0 & 3.4 & 0.0 & 4.0 \\
			& 0.2 & \textbf{22.4} & 6.4 & 5.6 & 0.0 & 9.6 \\
			& 0.4 & \textbf{89.6} & 19.8 & 4.6 & 0.0 & 39.6 \\
			& 0.6 & \textbf{100.0} & 58.4 & 6.4 & 0.0 & 93.8 \\
			& 0.8 & \textbf{100.0} & 94.2 & 6.6 & 0.0 & \textbf{100.0} \\
			\midrule
			\multirow{5}{*}{2000} & 0 & 5.8 & 4.6 & 4.8 & 0.0 & 5.0 \\
			& 0.2 & \textbf{9.0} & 4.8 & 5.2 & 0.0 & 5.0 \\
			& 0.4 & \textbf{21.8} & 5.2 & 6.6 & 0.0 & 6.2 \\
			& 0.6 & \textbf{50.0} & 12.0 & 4.6 & 0.0 & 12.6 \\
			& 0.8 & \textbf{80.6} & 21.2 & 4.2 & 0.0 & 18.4 \\
			\bottomrule
		\end{tabular}
	\end{table}
	
	\begin{table}[!htbp]
		\centering
		\caption{Empirical rejection rates (\%) for Example 2. The largest power is in bold.}
		\label{tab_exp2}
		\small
		\setlength{\tabcolsep}{20pt}
		\renewcommand{\arraystretch}{0.82}
		\begin{tabular}{ccrrr}
			\toprule
			$s$ & $p_{\mathrm{flip}}$ & GM$_1$ & NN$_{15}$ & MMD \\
			\midrule
			\multirow{5}{*}{7} & 0 & 5.0 & 4.6 & 5.4 \\
			& 0.2 & \textbf{12.0} & 4.8 & 5.6 \\
			& 0.4 & \textbf{34.2} & 11.0 & 7.2 \\
			& 0.6 & \textbf{83.0} & 54.0 & 10.0 \\
			& 0.8 & \textbf{100.0} & 98.6 & 21.0 \\
			\midrule
			\multirow{5}{*}{14} & 0 & 5.2 & 1.6 & 4.4 \\
			& 0.2 & \textbf{7.4} & 1.4 & 7.2 \\
			& 0.4 & \textbf{16.2} & 4.6 & 11.2 \\
			& 0.6 & \textbf{32.8} & 10.6 & 21.0 \\
			& 0.8 & \textbf{65.2} & 49.6 & 40.6 \\
			\midrule
			\multirow{5}{*}{21} & 0 & 6.2 & 0.0 & 6.2 \\
			& 0.2 & 6.6 & 0.8 & \textbf{7.0} \\
			& 0.4 & \textbf{14.0} & 0.6 & 12.8 \\
			& 0.6 & \textbf{29.6} & 3.2 & 27.4 \\
			& 0.8 & \textbf{63.8} & 17.2 & 60.6 \\
			\bottomrule
		\end{tabular}
	\end{table}

	For Example~1 (Table~\ref{tab_exp1}), GM$_1$, NN$_{15}$, MMD, and MCS have empirical sizes near the nominal $5\%$ level across the reported values of $p$. YLZ becomes increasingly conservative as $p$ grows, with zero observed rejections under the null for $p\in\{200,2000\}$.
	Under the alternative, GM$_1$ has lower power than some competitors when $p\in\{1,10\}$, while MCS is particularly strong in these lower-dimensional settings. For $p\in\{200,2000\}$, GM$_1$ has the largest power for all $p_{\mathrm{flip}}$ while YLZ never rejects. At $p=2000$, the power of NN$_{15}$, MMD, and MCS remains below $22\%$, $7\%$, and $19\%$, respectively, for all reported values of $p_{\mathrm{flip}}$. These results illustrate the difficulty of testing with many irrelevant conditioning coordinates in this simulation.
	
	For Example~2 (Table~\ref{tab_exp2}), our method and MMD maintain accurate size across all values of $s$, with sizes between 5.0\%-6.2\% and 4.4\%-6.2\% respectively. In contrast, NN$_{15}$ becomes increasingly undersized as $s$ grows, with size 1.6\% when $s=14$ and nearly zero when $s=21$.
	Under the alternative, our method has the highest reported power except at $s=21$ and $p_{\mathrm{flip}}=0.2$, where MMD is slightly higher. For NN$_{15}$, its power decreases drastically as $s$ increases and falls below 20\% 
	when $s=21$. In contrast, the power of MMD gets larger as $s$ increases, this is because, with fixed $p$, the proportion of components in $(X,Z)$ and $(Y,Z)$ that bear signal increases as $s$ increases, which makes two-sample tests more powerful. 
	
	\section{Imaging Data Applications}\label{sec:data}
	
	\begin{figure}[!htbp]
		\centering
		\begin{subfigure}[b]{0.47\textwidth}
			\centering
			\includegraphics[width=\textwidth]{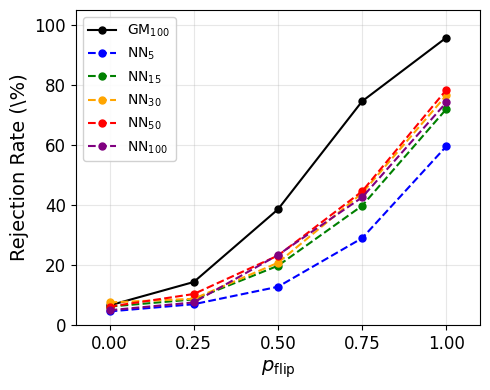}
			\caption{$n_T=500$}
			\label{fig:nT500}
		\end{subfigure}
		\hfill
		\begin{subfigure}[b]{0.47\textwidth}
			\centering
			\includegraphics[width=\textwidth]{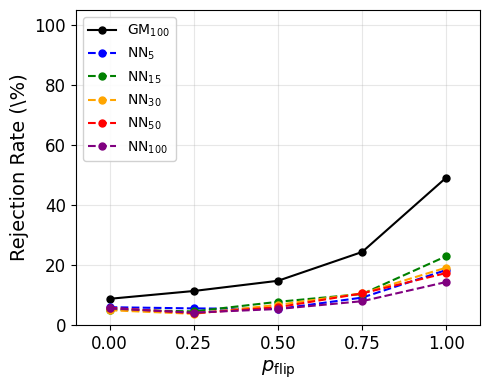}
			\caption{$n_T=3000$}
			\label{fig:nT3000}
		\end{subfigure}
		\par\smallskip
		\begin{subfigure}[b]{0.64\textwidth}
			\centering
			\includegraphics[width=\textwidth]{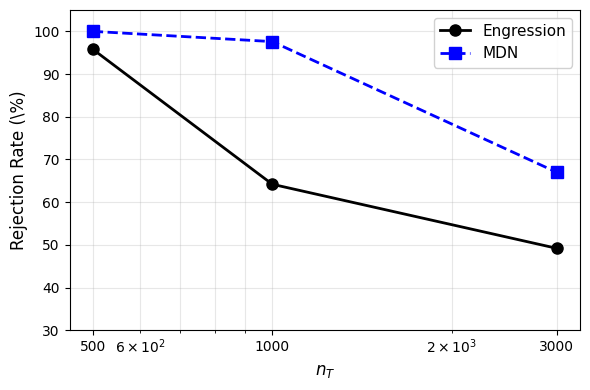}
			\caption{Engression vs MDN ($n_T$ in log scale)}
			\label{fig:engression_mdn}
		\end{subfigure}
		\caption{ Empirical rejection rates. (a) and (b): rejection rates as a function of $p_{\text{flip}}$ for $n_T=500$ and $n_T=3000$. (c): rejection rates of GM$_{100}$ at $p_{\text{flip}}=1$ for Engression and MDN as a function of $n_T$.}
		\label{fig:combined}
	\end{figure}
	
	In this section, we compare our proposed CDE test with the method of \citet{chatterjee2024ecmmd} for evaluating whether a conditional generative model can capture the conditional distribution of age given facial images. We also demonstrate how our test can be used to compare different conditional generative models for age prediction; see Section SM1 of the Supplementary Material for more implementation details.
	
	The dataset is the cropped and aligned UTKFace dataset\footnote{Available at \url{https://www.kaggle.com/datasets/abhikjha/utk-face-cropped}} \citep{zhang2017ageprogression}. After converting images to grayscale and restricting ages to 20--59, we obtain 16,425 image-label pairs. Ages are rescaled to $[0,1]$, and each image is processed using a pretrained ResNet-18 model \citep{he2016deep} to extract a 512-dimensional feature vector. The resulting dataset is denoted by $\{(W_i,Z_i)\}_{i=1}^{16425}$, where $W_i\in[0,1]$ is the rescaled age and $Z_i\in\R^{512}$ is the feature vector; see Section SM1.2 of the Supplementary Material for details.

	We split the full dataset of size $16{,}425$ into a training set of size $n_T$ and a test set of size $n_S = 16{,}425 - n_T$, with $n_T \in \{500, 1000, 3000\}$. Using the training set, we fit an Engression model $\widehat G_{n_T}$ to learn the conditional distribution of age given the feature vector. This fitted model is used to generate synthetic ages $Y_i$ for the evaluation and does not itself enter the test statistic; the generators in the statistic, which are used to generate $\wh Y_i^{(m)}$, are fitted separately using the test set. On the test set, for each feature vector $\{Z_i\}_{i=1}^{n_S}$, we generate a synthetic age $Y_i \sim \widehat G_{n_T}(\cdot \mid Z_i)$. 
	To introduce a controlled deviation from the null, for each $p_{flip} \in \{0, 0.25, 0.5, 0.75, 1\}$ we draw $\delta_i \stackrel{iid}{\sim} \mathrm{Bernoulli}(p_{flip})$ and define $X_i = W_i$ if $\delta_i = 1$, and otherwise set $X_i$ as an independent synthetic age drawn from $\widehat G_{n_T}(\cdot \mid Z_i)$. In this way, $X_i$ interpolates between the true response and model-generated samples. We then test the null hypothesis in equation~\eqref{eq_hypothesis} using $\{(Z_i, Y_i, X_i)\}_{i=1}^{n_S}$ to assess how well the \emph{fitted} $\widehat G_{n_T}$ approximates the true conditional distribution across different mixture levels. 
	
	Our test (denoted GM$_{100}$ with $M=100$) is applied 500 times on randomly selected subsamples of size $n=800$ drawn from $\{(X_i,Y_i,Z_i)\}_{i=1}^{n_S}$, and the empirical rejection rate at level $5\%$ is reported as a function of $p_{flip}$. For comparison, we include the test of \citet{chatterjee2024ecmmd} (denoted NN$_l$ for $l\in\{5,15,30,50,100\}$); see Section SM1.2 of the Supplementary Material for implementation details.
	As shown in Figure~\ref{fig:combined}, the rejection rate of GM$_{100}$ at $p_{flip}=1$ decreases as $n_T$ increases from 500 to 3000 (the analogous result for $n_T=1000$ is deferred to Figure SM1 in Section SM1.4 of the Supplementary Material). For $p_{flip}\ge 0.25$, GM$_{100}$ achieves substantially higher rejection rates than NN$_l$, indicating stronger power when the conditioning variable is high-dimensional. For $p_{flip}=0$, corresponding to $H_0$, both methods maintain rejection rates close to $5\%$. In addition, the rejection rates for NN$_l$ are sensitive to the choice of $l$, especially when $n_T$ is small, which may limit robustness in practical implementation.

	Finally, we compare different fitted generative models, such as Engression and mixture density networks (MDN), by replacing Engression with an MDN trained on the same training set and comparing rejection rates at $p_{flip}=1$ as a function of $n_T$. We report results only for our proposed test procedure, since it achieves the highest power in our simulation examples. As shown in Figure~\ref{fig:combined} (c), rejection rates are consistently higher when synthetic samples are generated from MDN than from Engression, indicating that Engression more accurately captures the conditional distribution when $Z$ is high-dimensional for this dataset, as its generated samples are less likely to be distinguished from the true responses than those produced by MDN. Moreover, as expected, the rejection rate decreases as $n_T$ increases because a larger training sample yields a more accurate estimate of the conditional distribution, making the generated samples closer to the true responses and harder to distinguish.

	\section{Conclusion}
	\label{sec:conclusion}
	
	We propose a paired conditional two-sample test based on ECMMD. Instead of estimating conditional mean embeddings by nearest-neighbor or kernel smoothing in the conditioning space, we estimate them through conditional sampling from learned generators. Sample splitting and studentization yield an asymptotically normal statistic under the null, provided the learned conditional mean embeddings are consistent. This avoids explicit smoothing over high-dimensional $Z$ and performs well empirically when the conditional generator can exploit low-dimensional or sparse structure in $Z$.

	From a theoretical perspective, we establish asymptotic normality of the studentized test statistic under the null and characterize its behavior under local alternatives converging to the null at polynomial rates. The test retains nontrivial power against alternatives shrinking at the $O(n^{-1/4})$ rate while remaining robust to estimation error in the learned conditional distributions. Notably, no convergence rate is required for estimating these conditional distributions under the null and fixed alternatives, allowing a broad class of generative models to be incorporated. Numerical studies further show that the method achieves accurate size and strong power in the high-dimensional setting.

	\section*{Supplementary material}
	The Supplementary Material contains implementation details, additional empirical results, and proofs of the lemma and theorems.

	\section*{Acknowledgements}
	Shao gratefully acknowledges partial support through a grant from the U.S. National Science Foundation.
	
	ChatGPT was used to assist in preparing the simulation code. The authors reviewed and validated the code and results. The authors assume responsibility for all content.

\bibliographystyle{biometrika}
\bibliography{CDE_testing}
\end{document}